\documentclass[a4paper]{article}
\usepackage{graphicx}
\usepackage{amsmath,amsthm,amssymb}

\usepackage{xcolor}

\usepackage{wrapfig}

\usepackage{caption}
\usepackage{subcaption}

\usepackage{svg}

\usepackage[utf8]{inputenc}

\usepackage[frozencache,cachedir=.]{minted}
\usepackage{algorithm}
\usepackage{algorithmic}

\usepackage[hidelinks]{hyperref}

\usepackage{paralist}

\usepackage{comment}

\usepackage{bbding}

\usepackage{indentfirst}

\usepackage{authblk}

\providecommand{\keywords}[1]
{
  \small	
  \textbf{Keywords---} #1
}

\title{Performance Comparison of Python Translators for a Multi-threaded CPU-bound Application}
\author[1]{Andr\'es Milla}
\author[2]{Enzo Rucci}

\affil[1]{Facultad de Inform\'atica,
Universidad Nacional de La Plata\\
La Plata, Buenos Aires, Argentina\\
 \authorcr
andressmilla@gmail.com}

\affil[2]{III-LIDI, Facultad de Inform\'atica,
Universidad Nacional de La Plata - CIC\\
La Plata, Buenos Aires, Argentina\\
 \authorcr
erucci@lidi.info.unlp.edu.ar}

\date{{May 23, 2022}}

\begin{document}

\maketitle              

\begin{center}
\texttt{This version of the contribution has been accepted for publication, after peer review (when applicable) but is not the Version of Record and does not reflect post-acceptance improvements, or any corrections. The Version of Record is available online at: \url{http://dx.doi.org/10.1007/978-3-031-05903-2\_2}. Use of this
Accepted Version is subject to the publisher’s Accepted Manuscript terms of use
\url{https://www.springernature.com/gp/open-research/policies/accepted-manuscript-terms}}
\end{center}

\begin{abstract}
Currently, Python is one of the most widely used languages in various application areas. However, it has limitations when it comes to optimizing and parallelizing applications due to the nature of its official CPython interpreter, especially for CPU-bound applications. To solve this problem, several alternative translators have emerged, each with a different approach and its own cost-performance ratio. Due to the absence of comparative studies, we have carried out a performance comparison of these translators using \textit{N-Body} as a case study (a well-known problem with high computational demand). The results obtained show that CPython and PyPy presented poor performance due to their limitations when it comes to parallelizing algorithms; while Numba and Cython achieved significantly higher performance, proving to be viable options to speed up numerical algorithms.

\end{abstract}

\keywords{Numba, Cython, N-body, CPU-bound, Parallel computing}

\clearpage

\section{Introduction}


Since it came out in the early 1990s, Python has now become one of the most popular languages~\cite{tiobe}. Still, Python is considered "slow" compared to compiled languages like C, C++, and Fortran, especially for CPU-bound applications~\footnote{Programs that perform a large number of calculations using the CPU exhaustively.}. Among the causes of its poor performance are its nature as an interpreted language and its limitations when implementing multi-threaded solutions~\cite{Marowka2018}. In particular, its main problem is the use of a component called \textit{Global Interpreter Lock} (GIL) in the official CPython interpreter. GIL only allows executing a single thread at a time, which leads to sequential execution. To overcome this limitation, processes are usually used instead of threads, but this comes at the cost of higher consumption of resources and a higher programming cost due to having a distributed address space~\cite{gil}. 

Even though there are alternative interpreters to CPython, some of them have the same problem, as in the case of PyPy~\cite{pypy_1}. In the opposite sense, some interpreters do not use GIL at all, such as Jython~\cite{jython_1}. Unfortunately, Jython uses a deprecated version of Python~\cite{python_deprecation,jython_1}, which limits future support for programs and the ability to take advantage of features provided by later versions of the language. Other translators allow the programmer to disable this component, as in the case of Numba, a JIT compiler that translates Python into optimized machine code ~\cite{numba_docs}.  Numba uses a Python feature known as decorators~\cite{decorators}, to interfere as little as possible in the programmer's code. Finally, Cython is a static compiler that allows transpiling~\footnote{Process performed by a special class of compiler, which consists in producing source code in one language based on source code in a different language.} Python codes to the equivalent C ones, and then compiling it to object code~\cite{cython_docs}.  It also allows disabling GIL and using C libraries, such as OpenMP~\cite{openmp}, which is extremely useful for developing multi-threaded programs.

When implementing a Python application, a translator must be selected. This choice is essential since it will not only impact program performance, but also the time required for development as well as future maintenance costs. To avoid making a “blind” decision, all relevant evidence should be reviewed. Unfortunately, the available literature on the subject is not exhaustive.

Even though there are studies that compare translators, they do so by using sequential versions~\cite{wilbers_using_nodate,roghult_benchmarking_nodate}, which does not allow assessing their parallel processing capabilities. On the contrary, if they consider parallelism, they do so between languages and not between Python translators~\cite{gmys_comparative_2020,wilk15,cai_performance_2005,varsha_review_nodate}. 

Based on the above, knowing the advantages and disadvantages of the different Python language translators is essential, both in sequential and multi-threaded contexts. This also applies to the primitives and functions that allow optimizing the code. Therefore, this article proposes a performance comparison between these, using the simulation of N computational bodies (\textit{N-Body}) - a CPU-bound problem that is popular in the HPC community - as case study. This paper is an extended and thoroughly revised version of~\cite{milla_rucci_cacic}. The work has been extended by providing:

\begin{itemize}
    \item An optimized implementation in the Cython language that computes N-Body on multicore architectures, which is
available in a public web repository for the benefit of the
academic and industrial communities~\footnote{\url{https://github.com/Pastorsin/python-hpc-study/}}.


    \item A comparative analysis of the performance of N-Body solutions  on a multicore architecture. This analysis can help Python programmers identify the strengths and weaknesses of each of them in a given situation.

\end{itemize}

The remaining sections of this article are organized as follows: in Section~\ref{sec:back}, the general background and other works that are related to this research are presented. Next, in Section~\ref{sec:imple}, the implementations used are described, and in Section~\ref{sec:results}, the experimental work carried out is detailed and the results obtained are analyzed. Finally, in Section~\ref{sec:conc}, our conclusions and possible lines of future work are presented.

\section{Background}
\label{sec:back}

\subsection{Numba}
\label{sec:numba}

Numba is a JIT compiler that allows translating Python code into optimized machine code using LLVM~\footnote{The LLVM Compiler Infrastructure, \url{https://llvm.org/}}. According to its documentation, Numba-based algorithms can approach the speeds of those of compiled languages like C, C++, and Fortran~\cite{numba_docs}, without having to rewrite its code thanks to an annotation-based approach called decorators~\cite{decorators}.

\begin{wrapfigure}{r}{0.4\textwidth}
    \centering
\begin{minted}{python}
from numba import njit

# Equivalent to indicating
# @jit(nopython=True)
@njit
def f(x, y):
    return x + y
\end{minted}
    \caption{Compilation in \textit{nopython} mode.}
    \label{fig:nopython}
\end{wrapfigure}

 \subsubsection{JIT compilation} 
 
 The library offers two compilation modes: (1) object mode, which allows compiling code that makes use of objects; (2) \textit{nopython} mode, which allows Numba to generate code without using the CPython API. To indicate these modes, the \texttt{@jit} and \texttt{@njit} decorators are used (see Fig.~\ref{fig:nopython}), respectively~\cite{numba_docs}.

By default, each function is compiled at the time it is called, and it is kept in cache for future calls. However, the inclusion of the \textit{signature} parameter will cause the function to be compiled at declaration time. In addition, this will also make it possible to indicate the types of data that the function will use and control the organization of data~\cite{numba_docs} in memory (see Fig.~\ref{fig:nopython_signatures}).
 
\subsubsection{Multi-threading} 

Numba allows enabling an automatic parallelization system by setting the parameter \texttt{parallel=True}, as well as indicating an explicit parallelization through the \texttt{prange} function (see Fig.~\ref{fig:nopython_parallel}), which distributes the iterations between the threads in a similar way to the OpenMP \texttt{parallel for} directive. It also supports reductions, and it is responsible for identifying the variables as private to each thread if they are within the scope of the parallel zone.
Unfortunately, Numba does not yet support primitives that allow controlling thread synchronization, such as semaphores or locks~\cite{numba_docs}.
    
\subsubsection{Vectorization}

Numba delegates code auto-vectorization and SIMD instructions generation to LLVM, but it allows the programmer to control certain parameters that could affect the task at hand, such as numerical precision using the \texttt{fastmath=True} argument. It also allows using \textit{Intel SVML} if it is available in the system~\cite{numba_docs}.

\subsubsection{Integration with NumPy}

It should be noted that Numba supports a large number of NumPy functions, which allows the programmer to control the memory organization of arrays and perform operations on them~\cite{numpy,numba_docs}.


\begin{figure}[t]
\centering
\begin{minipage}[b]{.49\textwidth}
    \centering
\begin{minted}{python}
from numba import njit, double

@njit(double(double[::1, :], 
                double[:, ::1]))
def f(x, y):
    """
    x:  Vector 2D of the "Double" type
        organized in columns.
    y:  Vector 2D of the "Double" type
        organized in rows.
    Returns the sum of the product 
    of vectors x and y.
    """
    return (x * y).sum()
\end{minted}
    \caption{Compilation in \textit{nopython} mode with the \mintinline{text}{signature} argument}
        \label{fig:nopython_signatures}
\end{minipage}\hfill%
\begin{minipage}[b]{.49\textwidth}
    \centering
\begin{minted}{python}
from numba import njit, double, prange

@njit(double(double[::1]), parallel=True)
def f(x):
    """
    x: 1D Vector.
    Returns the sum of vector x 
    through a reduction.
    """
    N = x.shape[0]    
    z = 0
    
    for i in prange(N):
        z += x[i]
        
    return z
\end{minted}
    \caption{Compilation in \textit{nopython} mode with the \mintinline{text}{parallel} argument}
    \label{fig:nopython_parallel}
\end{minipage}
\end{figure}

\subsection{Cython}

\label{subsec:cython}

Cython is a static compiler for Python created with the goal of writing C code taking advantage of the simple and clean syntax of Python~\cite{cython_docs}.
In other words, Cython is a Python superset that allows interacting with C functions, types, and libraries.

\subsubsection{Compilation} 
\begin{figure}[htp]
    \centering
\includegraphics[width=0.7\textwidth]{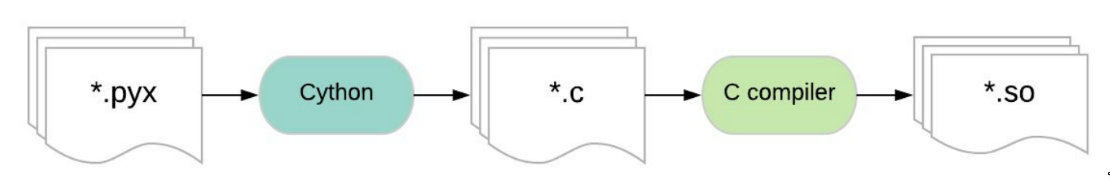} 
    \caption{Programming flow in Cython.}
    \label{fig:cython_flow}
\end{figure}

As shown in Fig.~\ref{fig:cython_flow} the Cython programming flow is very different from what the Python programmer is used to.

The main difference is that the file that will contain the source code has the extension \texttt{.pyx} unlike Python, where this extension \texttt{.py}. Then, this file can be compiled using a \texttt{setup.py} file, where compilation flags are provided to output:
\begin{inparaenum}[(1)]
    \item a file with \texttt{.c}, extension that corresponds to the code transpiled from Cython to C, and
    \item  a binary file with the extension \texttt{.so}, that corresponds to the compilation of the C file described previously. The latter will allow importing the compiled module into any Python script.
\end{inparaenum}

\subsubsection{Data types}

Cython allows declaring variables using C data types from the \texttt{cdef} statement (see Fig.~\ref{fig:cython_types_example}). While this is optional, it is recommended in the documentation to optimize program execution, since it avoids the inference of CPython types at runtime. In addition, Cython allows defining the memory organization for the arrays just like Numba~\cite{cython_docs}.


    
    

\subsubsection{Multi-threading}

Cython provides support for using OpenMP through the \texttt{cython.parallel}. This module contains the \texttt{prange}, function, which allows parallelizing loops using OpenMP's \texttt{parallel for} constructor. In turn, this function allows disabling GIL and defining the OpenMP \textit{scheduling} through the \texttt{nogil} and \texttt{schedule} arguments, respectively. 



It should be noted that all assignments declared within \texttt{prange} blocks are transpiled as \texttt{lastprivate}, while reductions are only identified if an \textit{in-situ} operator is used.  For example, the standard addition operation (\texttt{x = x + y}) will not be identified as a reduction, but the \textit{in-situ} addition operation (\texttt{x += y}), will (see Fig.~\ref{fig:cython_reduction}).



\subsubsection{Vectorization}
Cython delegates vectorization to the C compiler being used. Even though there are workarounds to force vectorization, it is not natively supported by Cython.

\subsubsection{Integration with NumPy}

Unfortunately, NumPy's vector operations are not supported by Cython. However, as mentioned above, NumPy can be used to control array memory organization.



\begin{figure}[htp]
\centering
\begin{minipage}[b]{.48\textwidth}
    \centering
\begin{minted}{cython}
cdef int x, y, z
cdef float a, b[100], *c

cdef struct Point:
    double x
    double y
\end{minted}
    \caption{Declared variables with C data types in Cython.}
    \label{fig:cython_types_example}
\end{minipage}\hfill%
\begin{minipage}[b]{.48\textwidth}
    \centering
\begin{minted}{cython}
from cython.parallel import prange

cdef int i
cdef int N = 30
cdef int total = 0

for i in prange(N, nogil=True):
    total += i
\end{minted}
    \caption{Reduction using Cython’s \texttt{prange} block.}
    \label{fig:cython_reduction}
\end{minipage}
\end{figure}

\subsection{The Gravitational N-body Simulation}

This problem consists in simulating the evolution of a system composed of N bodies during a time-lapse. Each body presents an initial state, given by its speed and position. The motion of the system is simulated through discrete instants of time. In each of them, every body experiences an acceleration that arises from the gravitational attraction of the rest, which affects its state.

\begin{figure}[hb!]

\begin{minted}{python}
for t in range(STEPS):
    for i in range(N):     
        for j in range(N):     
            Calculate the force exerted by C_j on C_i
            Sum the forces affecting C_i
        Calculate the displacement of  C_i
        Move C_i
\end{minted}
 \caption{Pseudo-code of the N-Body algorithm}
  \label{fig:nbody_pseudocode}
\end{figure}

The simulation is performed in 3 spatial dimensions and the gravitational attraction between two bodies $C_i$ and $C_j$ is computed according to Newtonian mechanics. Further information can be found at~\cite{nbody_ccis_cacic2020}.

The pseudo-code of the direct solution is shown in Fig.~\ref{fig:nbody_pseudocode}. This problem presents two data dependencies that can be noted in the pseudo-code. First, one body cannot move until the rest have finished calculating their interactions. Second, they cannot advance either to the next step until the others have completed the current step.
\section{N-Body Implementations}
\label{sec:imple}

In this section, the different implementations proposed are described.

\subsection{CPython Implementation}
\label{sec:cpython-imp}

\subsubsection{Naive implementation.}
\label{sec:python_naive}
Initially, a "pure" Python implementation (called \textit{naive}) was developed following the pseudocode shown in Fig.~\ref{fig:nbody_pseudocode},  which will serve as a reference to assess the improvements introduced later. It should be noted that this implementation uses Python lists as the data structure to store the state of the bodies.

\subsubsection{NumPy integration}

The use of NumPy arrays can speed up computation time since its core is implemented and optimized in C language. Therefore, it was decided to use these arrays as a data structure, further exploring the possible benefits of using the broadcasting function (operations between vectors provided by NumPy). Fig.~\ref{fig:python_numpy_broadcasting} presents the code of the CPython implementation with broadcasting.

\begin{figure}[t!]
    \centering
    \begin{minipage}{\textwidth}
\begin{minted}{python3}
def nbody(N, D, positions, masses, velocities, dp):
    # For each discrete instant of time
    for _ in range(D):
        # For every body that experiences a force
        for i in range(N):
            # Newton's Law of Universal Gravitation
            dpos = np.subtract(positions, positions[i])
            dsquared = (dpos ** 2.0).sum(axis=1) + SOFT
            gm = masses * (masses[i] * GRAVITY)
            d32 = dsquared ** -1.5
            gm_d32 = (gm * d32).reshape(N, 1)
            # Sum the forces
            forces = np.multiply(gm_d32, dpos).sum(axis=0)

            # Calculate acceleration of body i
            acceleration = forces / masses[i]
            # Calculate new velocity of body i
            velocities[i] += acceleration * DT / 2.0
            # Calculate new position of body i
            dp[i] = velocities[i] * DT

        # For every body that experienced a force
        for i in range(N):
            # Update position of body i
            positions[i] += dp[i]
\end{minted}
    \end{minipage}
    \caption{CPython implementation with \textit{broadcasting}.}
    \label{fig:python_numpy_broadcasting}
\end{figure}

\subsection{Numba Implementation}
\label{sec:numba-imp}

\subsubsection{Naive implementation}

The implementation described in Section~\ref{sec:python_naive}, which uses the operations between vectors provided by NumPy (\textit{broadcasting}), was selected as the initial implementation.

\subsubsection{Numba integration}

The first Numba version was obtained by adding a decorator to the \textit{naive} implementation (see lines 1-10 in Figure~\ref{fig:numba_parallel_calculate}).
The code was instructed to be compiled with relaxed precision using the \texttt{fastmath}, parameter, with NumPy's division model to avoid the divide-by-zero check (line 9)~\cite{numba_docs} and with \textit{Intel SVML}, which is inferred by Numba because it is available in the system.

\subsubsection{Multi-threading}

To introduce parallelism at the thread level, the \texttt{prange} statement was used. To do this, the loop that iterates over the bodies (line 5 in Figure~\ref{fig:python_numpy_broadcasting}) first had to be split into two parts. The first loop is responsible for computing Newton's law of gravitational attraction and Verlet's integration, while the second one updates body position.

\subsubsection{Arrays with simple data types}

NumPy's vector operations were replaced by numeric operations, and two-dimensional structures were replaced by one-dimensional ones to help Numba auto-vectorize the code (see Fig.~\ref{fig:numba_parallel}).

\begin{figure}[t!]
\centering
\begin{subfigure}{.5\textwidth}
 \centering
\begin{minted}{python}
@njit(
    void(
        int64,
        double[::1], double[::1], double[::1],
        double[::1],
        double[::1], double[::1], double[::1],
        double[::1], double[::1], double[::1],
    ),
    fastmath=True, parallel=True, error_model="numpy",
)
def calculate_positions(
    N,
    positions_x, positions_y, positions_z,
    masses,
    velocities_x, velocities_y, velocities_z,
    dp_x, dp_y, dp_z,
):
    # For every body that experiences a force
    for i in prange(N):
        # Initialize the force of the body i
        forces_x = forces_y = forces_z = 0.0
        # Calculate the forces exerted on body i
        # by every other body
        for j in range(N):
            # Newton's Law of Universal Gravitation
            dpos_x = positions_x[j] - positions_x[i]
            dpos_y = positions_y[j] - positions_y[i]
            dpos_z = positions_z[j] - positions_z[i]
            dsquared = (
                (dpos_x ** 2.0) + (dpos_y ** 2.0) +
                (dpos_z ** 2.0) + SOFT
            )
            gm = GRAVITY * masses[j] * masses[i]
            d32 = dsquared ** -1.5
            # Sum the forces
            forces_x += gm * d32 * dpos_x
            forces_y += gm * d32 * dpos_y
            forces_z += gm * d32 * dpos_z

        # Calculate acceleration of body i
        aceleration_x = forces_x / masses[i]
        aceleration_y = forces_y / masses[i]
        aceleration_z = forces_z / masses[i]
        # Calculate new velocity of body i
        velocities_x[i] += aceleration_x * DT / 2.0
        velocities_y[i] += aceleration_y * DT / 2.0
        velocities_z[i] += aceleration_z * DT / 2.0
        # Calculate new position of body i
        dp_x[i] = velocities_x[i] * DT
        dp_y[i] = velocities_y[i] * DT
        dp_z[i] = velocities_z[i] * DT
\end{minted}
   \caption{Function that calculates body positions.}
     \label{fig:numba_parallel_calculate}
\end{subfigure} \hfill
\begin{subfigure}{.49\textwidth}
  \centering
\begin{minted}{python}
@njit(
    void(
        int64,
        double[::1], double[::1], double[::1],
        double[::1], double[::1], double[::1],
    ),
    fastmath=True,
    parallel=True,
    error_model="numpy",
)
def update_positions(
    N,
    positions_x, positions_y, positions_z,
    dp_x, dp_y, dp_z,
):
    # For every body that experienced a force
    for i in prange(N):
        # Update position of body i
        positions_x[i] += dp_x[i]
        positions_y[i] += dp_y[i]
        positions_z[i] += dp_z[i]
\end{minted}
   \caption{Function that updates body positions.}
    \label{fig:numba_parallel_update}
\end{subfigure}
\caption{Numba implementation without \textit{broadcasting}}
\label{fig:numba_parallel}
\end{figure}

\subsubsection{Mathematical operations}
\label{sec:math}
The following alternatives for the calculation of the denominator of Newton's universal law of attraction are proposed: (1) calculating the positive power and then dividing; and (2) multiplying by the multiplicative inverse after calculating the positive power. Additionally, the following power functions are tested: (1) \texttt{pow} function in Python's math module; and (2) \texttt{power} function provided by NumPy.

\subsubsection{Vectorization}

As noted in Section~\ref{sec:numba}, Numba delegates auto-vectorization to LLVM. Even so, flags \texttt{avx512f, avx512dq, avx512cd, avx512bw, avx512vl} were defined to favor the use of this particular class of instructions.

\subsubsection{Data locality}
\label{sec:blocks}
To improve data locality, a version that iterates the bodies in blocks was implemented, similar to~\cite{Costanzo2021PerformanceVP}. 
To do this, the loop on line 19 in Figure ~\ref{fig:numba_parallel_calculate} will iterate over blocks of bodies, while in two other inner loops, Newton's gravitational attraction force and Verlet's integration will be calculated, respectively.

\subsubsection{Threading layer}
\label{numba_threading_layer}
Thread API changes were made through the \textit{threading layers} that Numba uses to translate parallel regions. To do this, different options were tested: \textit{default}, \textit{workqueue}, \textit{omp} (OpenMP) and \texttt{threading}. For the first three, the source code did not have to be modified, since Numba is responsible for translating the \texttt{prange} block to the selected API. However, this was not the case with \texttt{threading} – thread distribution had to be coded together with the specification of the parameter \texttt{nogil=True} to disable GIL. It should be noted that the \textit{tbb} option was not used because it was not available in the support server.

\subsection{Cython Implementation}
\label{sec:cython-imp}

\subsubsection{Naive implementation.}
\label{cython_naive}
As initial implementation (\textit{naive}), the one described in Section~\ref{sec:python_naive} was used. This implementation uses various NumPy arrays as data structures, which allows more flexible management of memory, particularly concerning organization and data types.

\subsubsection{Cython integration}
\label{cython_integracion}
No changes were made to the code of the \textit{naive} implementation (see Section~\ref{cython_naive}), but it was compiled using Cython. To do so, the extension of the code was simply changed from \texttt{.py} to \texttt{.pyx}.

\subsubsection{Explicit typing}
\label{cython_tipado}
As shown in Figure~\ref{fig:cython_parallel} the data types provided by Cython were explicitly defined to reduce interaction with the CPython API. Initially, to reduce unnecessary checks at runtime, the following compiler directives are provided (lines 1-4)~\cite{cython_docs}:

\begin{itemize}
    \item \textit{boundcheck} (line 1) avoids index error verifications on arrays.
    
    \item \textit{wraparound} (line 2) prevents arrays from being indexed relative to the end. For example, in Python, if A is an array with statement \texttt{A[-1]}, its last element can be obtained.
    
    \item \textit{nonecheck} (line 3) avoids verifications due to variables that can potentially take the value \textit{None}.
    
    \item \textit{cdivision} (line 4) performs the division through C avoiding CPython’s API. CPython.
\end{itemize}

On line 5, a hybrid function type is indicated through the \texttt{cpdef} statement, which allows the function to be imported from other applications developed in Python. Then, on lines 6-10, the Cython data types that will later be transpiled to C data types are specified. In particular, arrays are specified with the \texttt{double[::1]}, data type, which ensures that the arguments received are NumPy arrays contiguous in memory~\cite{cython_docs}. Finally, on lines 12-16, the data types corresponding to local variables are declared using the \texttt{cdef} statement.


\subsubsection{Multi-threading}
\label{cython_multihilado}
This version introduces thread-level parallelism through the \texttt{prange} statement provided by Cython. To do this, \texttt{range} statements were replaced by \texttt{prange} statements. In particular, the instruction was to use the \textit{static} policy as \textit{schedule} to evenly distribute the workload among the threads considering computation regularity. Additionally, GIL was disabled through the \textit{nogil} argument to allow these to be executed in parallel (see Fig.~\ref{fig:cython_parallel}).

Finally, it should be noted that the \texttt{prange} tatement is transpiled into an OpenMP \texttt{parallel for}~\cite{cython_docs}. Therefore, it has an implicit barrier that allows threads to be synchronized to comply with the data dependencies described in Figure~\ref{fig:nbody_pseudocode}.

\begin{figure}[t!]
\centering
\begin{minted}{cython}
@boundscheck(False)
@wraparound(False)
@nonecheck(False)
@cdivision(True)
cpdef void nbody(
    int N, int D, int T,
    double[::1] positions_x, double[::1] positions_y, double[::1] positions_z,
    double[::1] masses,
    double[::1] velocities_x, double[::1] velocities_y, double[::1] velocities_z,
    double[::1] dp_x, double[::1] dp_y, double[::1] dp_z,
):
    cdef double forces_x, forces_y, forces_z
    cdef double aceleration_x, aceleration_y, aceleration_z
    cdef double dpos_x, dpos_y, dpos_z
    cdef double dsquared, gm, d32
    cdef int i, j

    # For each discrete instant of time
    for _ in range(D):
        # For every body that experiences a force
        for i in prange(N, nogil=True, schedule="static", num_threads=T):
            # Initialize the force of the body i
            forces_x = forces_y = forces_z = 0.0

            # Calculate the forces exerted on body i by every other body
            for j in range(N):
                # Newton's Law of Universal Gravitation
                dpos_x = positions_x[j] - positions_x[i]
                dpos_y = positions_y[j] - positions_y[i]
                dpos_z = positions_z[j] - positions_z[i]
                dsquared = (dpos_x ** 2.0) + (dpos_y ** 2.0) 
                    + (dpos_z ** 2.0) + SOFT
                gm = GRAVITY * masses[j] * masses[i]
                d32 = dsquared ** -1.5
                # Sum the forces
                forces_x = forces_x + gm * d32 * dpos_x
                forces_y = forces_y + gm * d32 * dpos_y
                forces_z = forces_z + gm * d32 * dpos_z

            # Calculate acceleration of body i
            aceleration_x = forces_x / masses[i]
            aceleration_y = forces_y / masses[i]
            aceleration_z = forces_z / masses[i]
            # Calculate new velocity of body i
            velocities_x[i] += aceleration_x * DT / 2.0
            velocities_y[i] += aceleration_y * DT / 2.0
            velocities_z[i] += aceleration_z * DT / 2.0
            # Calculate new position of body i
            dp_x[i] = velocities_x[i] * DT
            dp_y[i] = velocities_y[i] * DT
            dp_z[i] = velocities_z[i] * DT

        # For every body that experienced a force
        for i in prange(N, nogil=True, schedule="static", num_threads=T):
            # Update position of body i
            positions_x[i] += dp_x[i]
            positions_y[i] += dp_y[i]
            positions_z[i] += dp_z[i]
\end{minted}
    
    \caption{Parallel Cython implementation.}
    \label{fig:cython_parallel}

\end{figure}

\subsubsection{Mathematical operations}
\label{cython_matematica}
A decision was made to evaluate the same alternatives for calculating the denominator for Newton's universal law of attraction as those described in Section~\ref{sec:numba-imp}. In particular, no changes were made to the power functions to avoid interaction with CPython’s API.

\subsubsection{Vectorization}
As mentioned in Section~\ref{subsec:cython},  Cython delegates auto-vectorization to the C compiler. However, the flags \texttt{-xCORE-AVX512, -qopt-zmm-usage=high, -march=native} were provided to favor the use of AVX-512 instructions.

\subsubsection{Data locality}
\label{cython_bloques}
To better take advantage of cache, a variant that iterates the bodies in blocks was implemented, similar to the one described in Section~\ref{sec:numba-imp}. In this version, the loop on line 21 in Figure~\ref{fig:cython_parallel} is split into two other inner loops. The first one calculates Newton's gravitational attraction force, while the second one calculates displacement using the \textit{velocity verlet} integration method.

\section{Experimental Results}
\label{sec:results}

\subsection{Experimental Design}

All tests were carried out on a Dell Poweredge server equipped with 2$\times$Intel Xeon Platinum 8276's with 28 cores (2 hw threads per core) and 256 GB of RAM. The operating system was Ubuntu 20.04.2 LTS, and the translators and libraries used were Python v3.8.10, PyPy v7.3.1, NumPy v1.20.1, Numba v0.52.0, Cython v0.29.22 and ICC v19.1.0.166.

For implementation evaluation, different workloads (\textit{N} = \{256, 512, 1024, 2048, 4096, 8192, 16384, 32768, 65536, 131072, 262144,524288\})
and number of threads (\textit{T} = \{1,56,112\}) were used. The number of simulation steps remained fixed (\textit{I}=100). Each proposed optimization was applied and evaluated incrementally based on the initial version~\footnote{Each previous version is labeled as \textit{Reference} in all graphics.}.
To evaluate performance, the GFLOPS (billion FLOPS) metric is used, with equation $GFLOPS=\frac{20 \times N^2 \times I}{t \times 10^9 }$, where \textit{N} is the number of bodies, \textit{I} is the number of steps, \textit{t} is execution time (in seconds) and factor 20 represents the number of floating-point operations required by each interaction~\footnote{A widely accepted convention in the literature for this problem.}.

\subsection{CPython Performance}

\begin{figure}[h!]

  \centering
  \includegraphics[width=.6\linewidth]{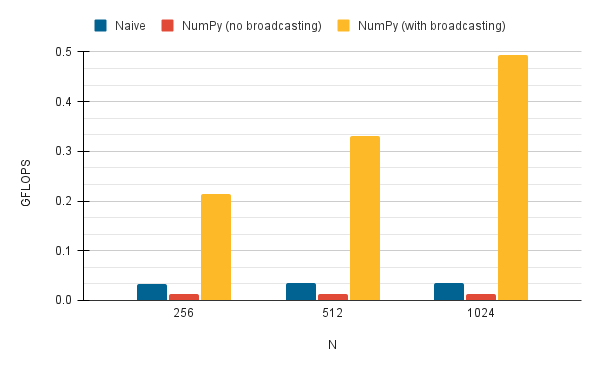}
  \captionof{figure}{CPython – Performance obtained with the different versions for various values of \textit{N}.}
  \label{fig:cpython}
\end{figure}

Fig.~\ref{fig:cpython} shows the performance obtained with the \textit{naive} version with NumPy when varying $N$. As it can be seen, the incorporation of NumPy arrays without using broadcasting worsened the performance by  ${2.9\times}$ on average. Because values are stored directly in NumPy arrays and must be converted to Python objects when accessed, unnecessary conversions are carried out using this algorithm. In contrast, this does not happen in the \textit{naive} version because values are already saved directly as Python objects.

This issue was solved by adding \textit{broadcasting}; that is, by performing vector operations between NumPy arrays. This avoids unnecessary conversions because operations are carried out internally in the NumPy core~\cite{numpy}. As it can be seen, performance improved by ${10\times}$ on average with respect to the \textit{naive} version.

\begin{figure}[h!]

  \centering
  \includegraphics[width=.6\linewidth]{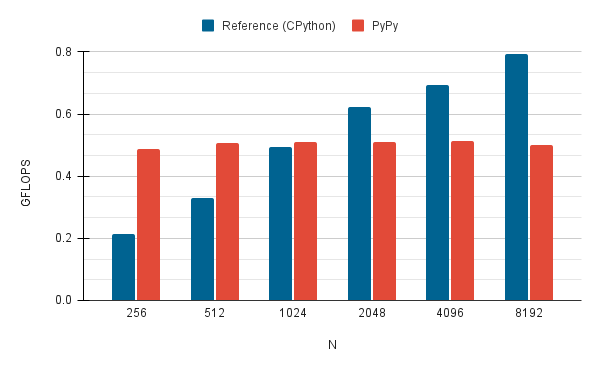}
  \captionof{figure}{PyPy – Performance obtained for various values of \textit{N}.}
  \label{fig:pypy}
\end{figure}

Finally, Fig.~\ref{fig:pypy} shows the performance of CPython compared to PyPy. As it can be seen, CPython's performance tends to improve as size increases, whereas PyPy's~\footnote{The implementation executed with PyPy was the naive version since the other versions use NumPy arrays and PyPy is unable to optimize them.} performance remains constant.

\clearpage
\subsection{Numba Performance}

\begin{figure}[h!]

  \centering
  \includegraphics[width=.6\linewidth]{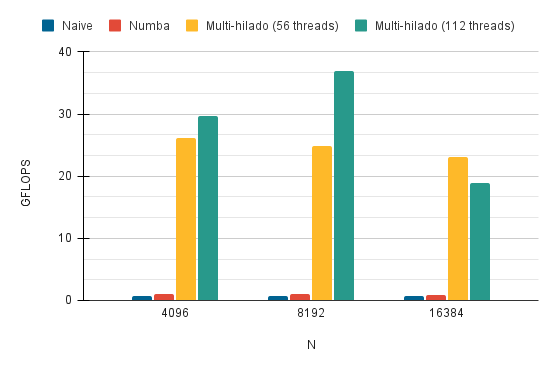}
  \captionof{figure}{Numba – Performance obtained for compiling and multi-threading options for various values of \textit{N}.}
  \label{fig:compiler}
\end{figure}

Fig.~\ref{fig:compiler} shows the performance when activating the compilation options and applying multi-threading with various values of \textit{N}. Even though the Numba compilation options (\texttt{njit+fastmath+svml}) have practically no effect on the performance of this version, there is a significant improvement when using threads to compute the problem. In particular, an average improvement of 33$\times$ and 38$\times$ for 56 and 112 threads, respectively, can be noted. 

\begin{figure}[h!]

  \centering
  \includegraphics[width=.6\linewidth]{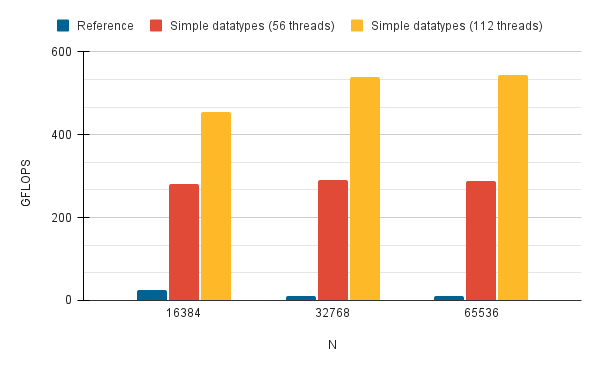}
  \captionof{figure}{Numba – Performance obtained with the different versions for various values of \textit{N}.}
  \label{fig:parallel}
\end{figure}

Fig.~\ref{fig:parallel} shows the significant improvement obtained when using arrays with simple data types instead of compound ones (an average of 41$\times$ in the case of 112 threads). Even though the second one simplifies coding, it also involves organizing the data in the form of an \textit{array of structures}, which imposes limitations on the use of the SIMD capabilities of the processor ~\cite{Intel_aos_soa}. Additionally, it can also be noted that using \textit{hyper-threading}  results in an improvement of approximately 78\% in this case.

\begin{figure}[h!]

  \centering
  \includegraphics[width=.6\linewidth]{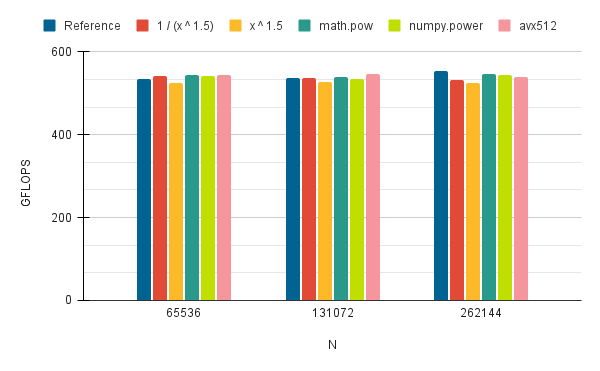}
  \captionof{figure}{Numba – Performance obtained using different mathematical calculations, power functions and AVX512 instructions for various values of \textit{N}.}
  \label{fig:math}
\end{figure}

There are practically no changes in performance if the mathematical calculations and power functions used are different from those described in Section~\ref{sec:numba-imp} (see Fig.~\ref{fig:math}). This is because no matter which option is used, the resulting machine code is always the same. Something similar happens when explicitly specifying the use of AVX-512 instructions. As mentioned in Section~\ref{sec:numba}, Numba attempts to auto-vectorize the code via LLVM. Looking at the machine code, it was observed that the generated instructions already made use of these extensions.

\begin{figure}[h!]

  \centering
  \includegraphics[width=.6\linewidth]{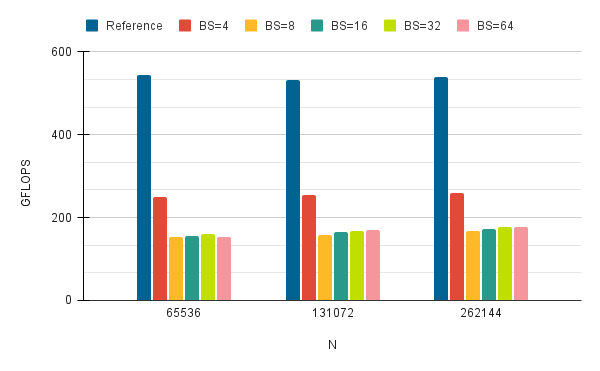}
  \captionof{figure}{Numba – Performance obtained when processing in blocks for various values of \textit{N}.}
  \label{fig:blocks}
\end{figure}

The processing by blocks described in Section~\ref{sec:numba-imp} did not improve the performance of the solution, as it can be seen in Fig.~\ref{fig:blocks}. The performance loss is related to the fact that this computation reorganization produces failures in LLVM when auto-vectorizing. Unfortunately, since Numba does not offer primitives to specify the use of SIMD instructions explicitly, there is no way to fix this.

\begin{figure}[h!]

  \centering
  \includegraphics[width=.6\linewidth]{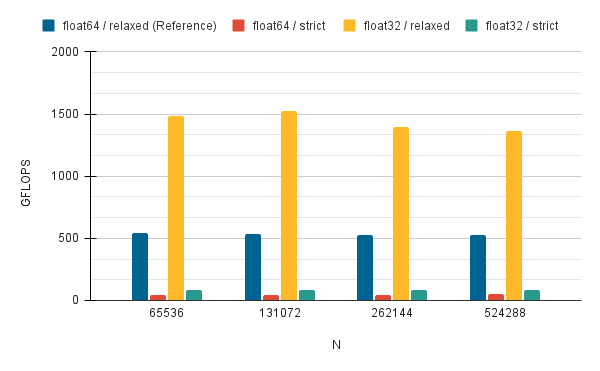}
  \captionof{figure}{Numba – Performance obtained for precision reduction for various data types and values of \textit{N}.}
  \label{fig:datatype}
\end{figure}

Fig.~\ref{fig:datatype} shows the performance obtained for precision reduction with various data types and workloads (\textit{N}). It can be seen that the use of the \texttt{float32} data type (instead of \texttt{float64}) leads to an improvement of up to 2.8$\times$ GFLOPS, at the cost of a reduction in precision. Similarly, relaxed precision produced a significant acceleration  on both data types: \texttt{float32} (17.2$\times$ on average) and \texttt{float64} (11.4$\times$ on average). In particular, the performance peak is 1524/536 GFLOPS in single/double precision. It is important to mention that this version achieves an acceleration of 687$\times$ compared to the \textit{naive} implementation (\texttt{float64}).

\begin{figure}[h!]

  \centering
  \includegraphics[width=.6\linewidth]{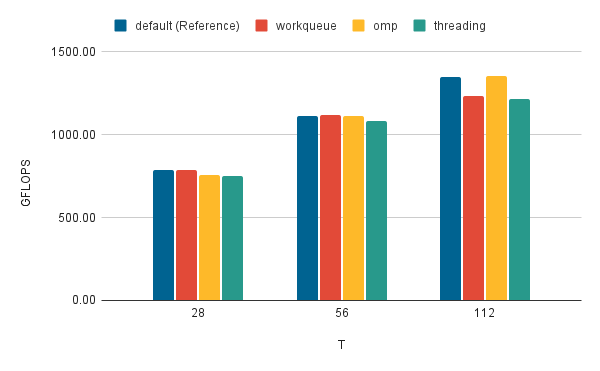}
  \captionof{figure}{Numba – Performance obtained with the different \textit{threading layers} in Numba for various values of \textit{T} and a fixed value of ${N=524288}$.}
  \label{fig:numba_threading_layer}
\end{figure}

Finally, Fig.~\ref{fig:numba_threading_layer} shows that, when using 112 threads, OpenMP and the default \textit{threading layer} provided by Numba outperform the others by an average of ${9.3\%}$.
In turn, it can be seen that the use of \texttt{threading} is below all tested threading layers. This is due to the overhead generated by the use of Python objects to synchronize the threads, which does not occur in the other threading layers because synchronization is carried out on its own API.

\clearpage
\subsection{Cython Performance}

\begin{figure}[h!]

  \centering
  \includegraphics[width=.6\linewidth]{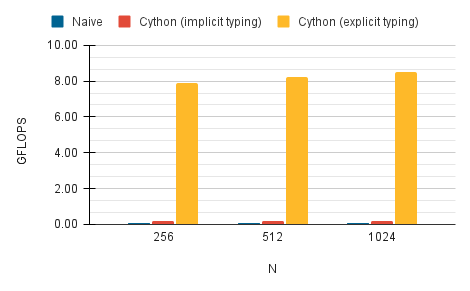}
  \captionof{figure}{Cython – Performance obtained with and without explicit typing for various values of \textit{N}.}
  \label{fig:cython_types_chart}
\end{figure}

Fig.~\ref{fig:cython_types_chart} shows that the Cython integration (without specifying the data types for the variables) did not produce a significant improvement concerning the \textit{naive} version, since the resulting code is using CPython’s API. 
On the other hand, defining the variables with Cython data types reduces this interaction and, therefore, performance improves remarkably (${547.7\times}$ on average).

\begin{figure}[h!]

  \centering
  \includegraphics[width=.6\linewidth]{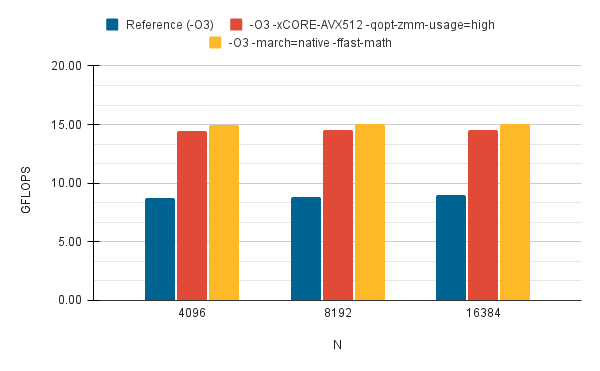}
  \captionof{figure}{Cython – Performance obtained with the different compilation options for various values of \textit{N}.}
  \label{fig:cython_compilation_chart}
\end{figure}

As shown in Fig.~\ref{fig:cython_compilation_chart}, specifying the use of AVX-512 instructions achieved an improvement of ${1.7\times}$ on average. In particular, the performance obtained with the \texttt{-march=native} flag was slightly higher (${1.4}$ GFLOPS on average) than that obtained with the \texttt{-xCORE-AVX512 -qopt-zmm-usage=high} flag.

\begin{figure}[h!]

  \centering
  \includegraphics[width=.6\linewidth]{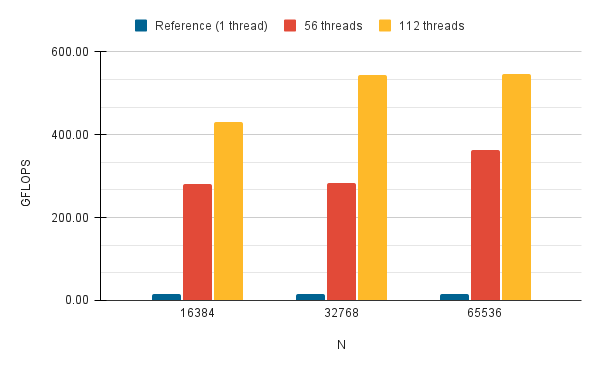}
  \captionof{figure}{Cython – Performance obtained with the multi-threaded solution for various values of \textit{N}.}
  \label{fig:cython_parallel_chart}
\end{figure}

\begin{figure}[h!]

  \centering
  \includegraphics[width=.6\linewidth]{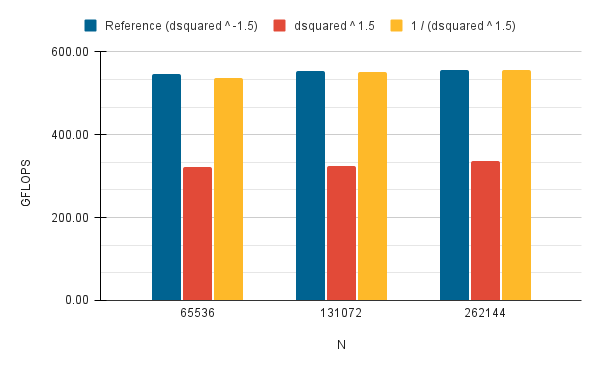}
  \captionof{figure}{Cython – Performance obtained with the mathematical calculations for various values of \textit{N}.}
  \label{fig:cython_math_chart}
\end{figure}

Both multi-threading and mathematical optimizations led to positive results. On the one hand, Fig.~\ref{fig:cython_parallel_chart} shows that the multi-threaded solution with 56 and 112 threads achieved a remarkable improvement of ${21.1\times}$ and ${34.6\times}$ on average, respectively. On the other hand, Fig.~\ref{fig:cython_math_chart} shows the performance obtained when applying the mathematical operations described in Section~\ref{sec:cython-imp}. As it can be seen, using a direct division degraded the performance by ${41\%}$; while calculating the multiplicative inverse by positive power did not result in any significant improvements.

\begin{figure}[h!]

  \centering
  \includegraphics[width=.6\linewidth]{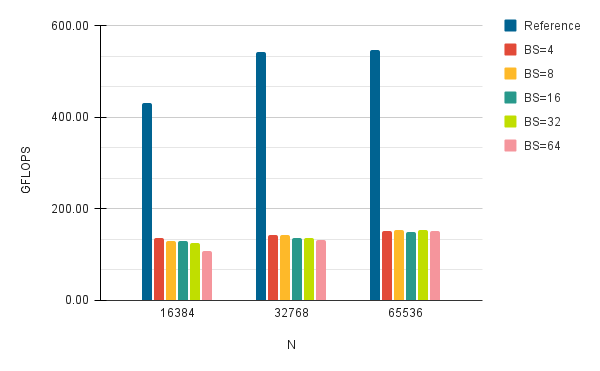}
  \captionof{figure}{Cython – Performance obtained when processing in blocks for various values of \textit{N}.}
  \label{fig:cython_blocking_chart}
\end{figure}

Block processing significantly worsened solution performance for all tested block sizes (see Fig.~\ref{fig:cython_blocking_chart}). This is because the compiler identifies false dependencies in the code and does not generate the corresponding SIMD instructions. Unfortunately, this cannot be fixed, as Cython does not provide a way to tell the compiler that it is safe to vectorize operations.

\begin{figure}[h!]

  \centering
  \includegraphics[width=.6\linewidth]{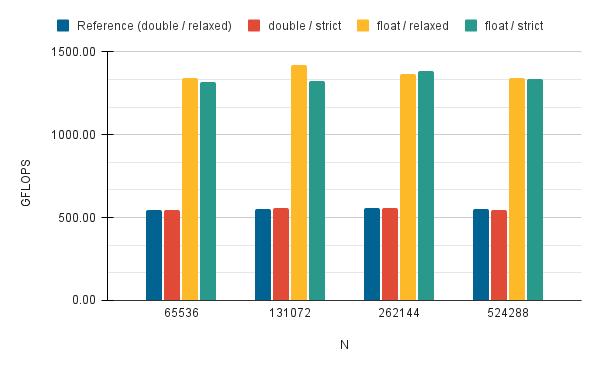}
  \captionof{figure}{Performance obtained for precision reduction for various data types and values of \textit{N}.}
  \label{fig:cython_precision_chart}
\end{figure}

Finally, Fig.~\ref{fig:cython_precision_chart} shows that precision reduction had practically no effect on the performance obtained. However, using \textit{float} as data type improved performance noticeably (${1362}$ GFLOPS on average) at the cost of less representation in the final result.

\clearpage
\subsection{Performance Comparison}

\begin{figure}[h!]

  \centering
  \includegraphics[width=.6\linewidth]{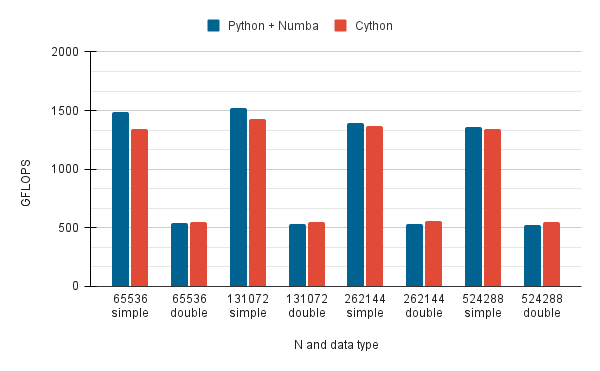}
  \captionof{figure}{Performance comparison between the final versions of Numba and Cython for various data types and values of \textit{N}.}
  \label{fig:cython_vs_numba_chart}
\end{figure}

First of all, it should be noted that the versions with CPython and PyPy were not included in the final comparison due to their low performance ($0.5$ GFLOPS on average). Fig.~\ref{fig:cython_vs_numba_chart} shows a comparison between the optimized implementations of Numba and Cython for various workloads and data types. As it can be seen, when using double precision, Cython was slightly faster than Numba by an average of $16.7$ GFLOPS; while in single precision, Numba was superior by an average of $73$ GFLOPS. These values represent improvements of $3\%$ and $5\%$, respectively. In turn, it is important to mention that both final versions of Numba and Cython achieved an average acceleration of $1018\times$ and $1050\times$ respectively, compared to the best CPython implementation (\texttt{float64}).

\section{Conclusions and Future Work}
\label{sec:conc}



In this work, a performance comparison has been made between CPython, PyPy, Numba, and Cython. In particular, N-Body -a parallelizable problem with high computational demand and considered to be  CPU-bound- was chosen as a case study. To this end, different algorithms were produced for each translator, starting from a base version and applying incremental optimizations until reaching the final version. In this sense, the benefits of using multi-threading, block processing, \textit{broadcasting}, different mathematical calculations and power functions, vectorization, explicit typing, and different thread APIs were explored.

Considering the results obtained, it can be said that there were no significant differences between the performance of Numba and Cython. However, both translators significantly improved CPython's performance. This was not the case with PyPy, since it failed to improve the performance of CPython+NumPy due to its inability to parallelize, associated with GIL.
Therefore, it can be stated that in contexts similar to those of this study, both Numba and Cython can be powerful tools to accelerate CPU-bound applications developed in Python. The choice between one or the other will be largely determined by the approach that the development team finds most convenient, considering the characteristics of each one.

As future work, it would be interesting to extend on the following directions:
\begin{itemize}

    \item Replicating the study carried out considering: \begin{inparaenum}[(1)]
        \item other case studies that are computationally intensive but whose characteristics are different from those of \textit{N-Body};
        \item other multicore architectures different from the one used in this work.
    \end{inparaenum} 
    
    
    \item Considering that programming effort is an increasingly relevant issue~\cite{progamming_effort_microsoft_article}, comparing the solutions developed from this perspective.

    \item Given that other technologies allow parallelism to be implemented at the process level in Python, comparing these considering not only their performance but also programming cost.
\end{itemize}

%
%
%
\bibliographystyle{splncs04}
\bibliography{references}
\end{document}